# Influence of doping on non-equilibrium carrier dynamics in graphene

*Leonard Weigl[1], Johannes Gradl[1], Peter Richter[2], Thomas Seyller[2], Domenica Convertino[3], Stiven Forti[3], Camilla Coletti[3], Isabella Gierz[1]\**

[1]Institute for Experimental and Applied Physics, University of Regensburg, 93040 Regensburg, Germany

[2] Institute for Physics, Technical University Chemnitz, 09126 Chemnitz, Germany

[3]Center for Nanotechnology Innovation@NEST, Istituto Italiano di Tecnologia, 56127 Pisa, Italy

ABSTRACT: Controlling doping is key to optimizing graphene for high-speed electronic and optoelectronic devices. However, its impact on non-equilibrium carrier lifetimes remains debated. Here, we systematically tune the doping level of quasi-freestanding epitaxial graphene on SiC(0001) via potassium deposition and probe its ultrafast carrier dynamics directly in the band structure using time- and angle-resolved photoemission spectroscopy (trARPES). We find that increased doping lowers both the peak electronic temperature and the cooling rate of Dirac carriers, which we attribute to higher electronic heat capacity and reduced phonon emission phase space.

Comparing quasi-freestanding graphene with graphene on a carbon buffer layer reveals faster relaxation in the latter, likely due to additional phonon modes being available for heat dissipation. These findings offer new insights for optimizing graphene in electronic and photonic technologies.

INTRODUCTION: Doping plays a key role in controlling charge carriers in modern electronic devices. As the demand for faster device operation and higher clock rates continues to grow, understanding the effects of doping on non-equilibrium carrier dynamics becomes increasingly important. Its ultrahigh carrier mobility allowing for ballistic transport at room temperature [1, 2] together with its high thermal conductivity [3], mechanical flexibility [4] and broadband frequency response [5] make graphene an ideal material for future high-speed electronic and optoelectronic devices. However, the impact of doping on the non-equilibrium carrier dynamics of graphene remains a topic of debate in the literature. Some studies [6, 7] report that higher doping levels lead to faster relaxation, while others [8] suggest that carrier decay times are independent of doping. This highlights the need for further experimental and theoretical investigations to fully understand and control graphene's carrier dynamics.

Here, we systematically tune the doping level of quasi-freestanding epitaxial graphene on SiC(0001) [9] via potassium deposition and directly probe the non-equilibrium carrier dynamics within the band structure using time- and angle-resolved photoemission spectroscopy (trARPES). Our results reveal that both the peak electronic temperature of the Dirac carriers and their cooling rate decrease with increasing doping. We attribute these trends to the rise in electronic heat capacity with doping and the combined influence of the density of states at the Fermi level and electronic temperature on the scattering phase space available for phonon emission.

Furthermore, we compare the non-equilibrium carrier dynamics of quasi-freestanding graphene [9] with the one of graphene resting on a carbon buffer layer [10]. We find that carrier relaxation in graphene on a carbon buffer layer is consistently faster than in quasi-freestanding graphene across all doping levels studied. This suggests that a previous interpretation based solely on differences in scattering phase space [6] is too simplistic. Instead, we propose that the observed differences between these two graphene systems may be linked to the distinct phonon modes available for heat dissipation.

Our study clarifies how doping and different substrates influence carrier dynamics in graphene, leading to improved strategies for optimizing the performance of graphene in various electronic and photonic applications.

METHODS: sample preparation: N-doped, single-side polished 6H-SiC(0001) substrates were purchased from SiCrystal. The substrates were cleaned with acetone and IPA in an ultrasonic bath and a subsequent HF treatment. Next, the substrates were H-etched at 1200 °C in molecular hydrogen for 5-10 minutes to remove polishing scratches [11] and graphitized in Ar atmosphere [10] in an Aixtron high temperature BM cold-wall reactor. Graphitization at 1200 °C resulted in the formation of a carbon-rich buffer layer with $(6\sqrt{3}\times6\sqrt{3})R30°$ reconstruction. Further graphitization at 1300-1350 °C for 5-10 minutes resulted in the formation of a graphene monolayer on top of the carbon-rich buffer layer (G/C-SiC). Alternatively, quasi-freestanding monolayer graphene (G/H-SiC) was grown on on-axis oriented Si-face n-type SiC with an epi-ready chemo-mechanical polish purchased from Cree, which was thoroughly degreased. Next, a carbon-rich buffer layer with $(6\sqrt{3}\times6\sqrt{3})R30°$ periodicity was formed via polymer-assisted sublimation growth (PASG) [12], which included deposition of an ultrathin layer of AZ5214E photoresist by spin coating, followed by three subsequent annealing steps at 900°C for 30 minutes in vacuum and at 1200°C (1400°C) for 12 minutes (6 minutes) in 1000 mbar of Ar, using a custom built induction furnace [13]. G/H-SiC was then obtained by intercalation of hydrogen [9] which was performed by annealing the sample in 860 mbar ultra-pure hydrogen at 550°C for 90 minutes [14].

trARPES: The setup is based on a Ti:Sa amplifier with a repetition rate of 1kHz, a pulse energy of 7mJ, and a pulse duration of 35fs (Astrella, Coherent). The samples were photoexcited with a pump photon energy of 1.55eV and a pump fluence of 0.9mJ cm$^{-2}$. To generate extreme ultraviolet (XUV) probe pulses, the output of the Ti:Sa amplifier was frequency doubled and focused onto an Argon gas jet for high harmonics generation. A single harmonic at 21.7eV was then selected with a grating monochromator. ARPES spectra were recorded with a hemispherical analyser (Phoibos 100, Specs). The energy and temporal resolution of the experiments was ~ 200meV and ~ 250fs, respectively.

RESULTS: Figure 1a shows ARPES spectra of the different graphene samples taken with a non-monochromatized He lamp at the K point along the ΓK direction of the hexagonal Brillouin zone. Orange lines are tight-binding fits of the electronic dispersion. Note that due to many-body interactions a linear extrapolation of the valence band does not pass through the conduction band [15] which is why we include an artificial band gap in Figs. 1 a4 and a5. The position of the Dirac point is assumed to be in the centre of this artificial gap. The Dirac point of pristine G/H-SiC (Fig. 1 a1) is found to be $E_D \approx$ 100meV above the Fermi level consistent with literature [9]. Upon potassium deposition (Figs. 1 a2-a4) the Dirac cone is found to move to lower energies. The maximum doping level for G/H-SiC considered in the present study is $E_D \approx$ -300meV in Fig. 1 a4. Here, a second Dirac cone (light red in Fig. 1 a4) starts to emerge, possibly indicating the onset of potassium intercalation [16]. The doping level of pristine G/C-SiC with the Dirac point $E_D \approx$ -370meV below the Fermi level (Fig. 1 a5) in good agreement with literature [10, 15].

Figure 1b shows corresponding ARPES spectra measured with femtosecond XUV pulses at negative pump-probe delay before the arrival of the pump pulse. Here, only one of the two branches of the Dirac cone is visible due to photoemission matrix element effects for our p-polarized probe pulses [17]. The pump-induced changes of the photocurrent with respect to negative pump-probe delays are shown in Fig. 1c for a pump-probe delay of ~ 140fs. Red and blue indicate gain and loss of photoelectrons, respectively. All spectra show a gain above the Fermi level and a corresponding loss below the Fermi level indicating a redistribution of the carriers by the pump pulse.

To analyse this pump-probe signal in detail we start by integrating the spectra in Fig. 1b over the full momentum axis (Figs. 1 b1-b3 and b5) and over the momentum range indicated by the blue-shaded area in Fig. 1 b4, respectively. The resulting energy distributions curves (see Fig. 2a for G/H-SiC) are fitted with a Fermi-Dirac distribution convolved with a Gaussian accounting for the finite energy resolution. The fit also considers the density of states of graphene in the momentum range defined by the width of the analyser slit. These fits directly yield the transient electronic temperature of the carriers inside the Dirac cone displayed in Fig. 2b for different doping levels as indicated in the figure. The temporal evolution of the electronic temperature is fitted with a single exponential decay (continuous lines in Fig. 2b). Note that hot carrier dynamics in graphene typically follow a double exponential decay with decay times of about hundred femtoseconds and some picoseconds due to optical and acoustic phonon emission, respectively [18, 19, 20, 21, 22]. In the present study, however, the temporal resolution was insufficient to properly resolve these distinct timescales. The peak electronic temperature is found to decrease with increasing doping level $|E_D|$ (Fig. 2c). Further, Fig. 2d shows that the exponential cooling time for the Dirac electrons increases with increasing doping level for potassium doped G/H-SiC. The cooling time for G/C-SiC is found to be 530±140fs, ~ 620fs lower than the cooling time extrapolated for K-doped G/H-SiC with a similar doping level.

Further information about carrier cooling via optical phonon emission can be obtained by analysing the energy dependence of the population lifetimes of the carriers inside the Dirac cone [23, 24]. To gain access to this information, we integrate the counts over the areas marked by the coloured boxes in Fig. 1 c1 yielding the pump-probe traces in Fig. 3a. Single-exponential fits (continuous lines in Fig. 3a) then provide the energy-resolved population lifetimes shown in Fig. 3b. For all samples the decay times are found to increase when approaching the Fermi level consistent with literature [6, 23, 24, 25]. For K-doped G/H-SiC, within the experimental error bars, the energy-resolved lifetimes are found to be independent of the doping level. The values for G/C-SiC are found to be a factor of ~2 lower compared to K-doped G/H-SiC.

In summary, we find that (1) the peak electronic temperature decreases with increasing doping level, (2) for K-doped G/H-SiC the exponential cooling time of the electrons increases with increasing doping level, and (3) carrier relaxation in G/C-SiC is faster than in K-doped G/H-SiC for all doping levels investigated in the present study.

DISCUSSION: (1) Our findings are in good agreement with previous trARPES data [6] that reported peak electronic temperatures of ~3300K and ~2000K for G/H-SiC and G/C-SiC, respectively, for a pump fluence of 1.4mJ cm$^{-2}$. According to standard text books [26] the electronic heat capacity can be approximated by $C_e = D(E_F) T_e k_B^2 \pi^2/3$, where $D(E_F)$ is the electronic density of states at the Fermi level. In graphene, $D(E) \propto |E|$ and $C_e$ increases linearly with $E_F$. Hence, the peak electronic temperature is expected to decrease with increasing carrier concentration as observed in Fig. 2c. Note that, in order to make quantitative predictions for the peak electronic temperature for different doping levels, much more complex theories need to be applied that consider absorption saturation due to Pauli blocking as well as an ultrafast redistribution of photoexcited electron-hole pairs on timescales comparable to typical pump and probe pulse durations.

(2) The time scale for carrier cooling is determined by the phase space available for phonon emission that depends on the electronic density of states within an energy interval of $\pm k_B T_e$ around the Fermi level (see sketch in Fig. 4). While doping increases the density of states at the Fermi level, it also reduces the peak electronic temperature. We find that the cooling rate $1/\tau$ decreases with increasing doping level (Fig. 3b). This indicates that the influence of the increased density of states is overcompensated by the decreased electronic temperature leading to an overall reduction of scattering phase space. Our observation contradicts previous results from optical-pump terahertz-probe spectroscopy that either show shorter relaxation times for higher doping levels [7] or decay times of ~ 2ps independent of doping level [8]. Although we cannot explain these differences, we would like to point out that the experiments in [7, 8] were done with pump fluences that were one to two orders of magnitude lower than in the present experiment. Further, [7, 8] explored the differential transmission in the terahertz spectral range corresponding to an energy window of ± 10meV around the Fermi level, while we probed the non-equilibrium carrier dynamics over a much broader frequency range of ~ 1eV. Finally, the trARPES studies in the present manuscript provide direct access to the transient electronic temperature, while in [7] the electronic temperature was obtained from phenomenological fits of the Drude scattering rate that itself was obtained by extracting the Drude conductivity from the experimentally measured differential transmission.

(3) Also here, our observation of faster carrier relaxation for G/C-SiC compared to G/H-SiC (Fig. 2d and Fig. 3b) is in good agreement with previous trARPES results from [6]. The results in [6] were attributed to the scattering phase space for phonon emission that was believed to be bigger for G/C-SiC than for G/H-SiC due to their differences in doping level. According to [6], we would expect similar cooling times for G/C-SiC and K-doped G/H-SiC provided that the Fermi levels are similar. Our results in Fig. 2d, however, show that carrier cooling in G/C-SiC is always faster than in G/H-SiC independent of the doping level of G/H-SiC. This indicates that the previous interpretation [6] was too simple. Instead, we tentatively attribute the increased cooling rate for G/C-SiC to the availability of additional phonon modes for heat dissipation. One possible candidate are phonon modes of the 6√3 carbon buffer layer. Alternatively, it is also conceivable that an increased defect density of G/C-SiC increases the coupling between hot electrons and acoustic phonons of the graphene layer itself [27, 28]. In principle, each phonon mode should leave its fingerprints in the energy-dependent population lifetimes plotted in Fig. 3b. However, both the finite energy resolution and the limited signal-to-noise ratio of our data do not allow us to distinguish the contribution of different phonon modes.

SUMMARY AND OUTLOOK: We investigated the carrier dynamics of potassium-doped quasi-freestanding epitaxial graphene on SiC(0001) with time- and angle-resolved photoemission spectroscopy. We find that both the peak electronic temperature and the cooling rate of the Dirac carriers decrease with increasing doping level. We interpret these observations in terms of an increased electronic heat capacity and a reduced phase space for phonon emission. We also explored the role of different substrates and found that carrier relaxation is consistently faster in G/C-SiC than in potassium-doped G/H-SiC, challenging prior interpretations based solely on doping differences. This suggests that additional phonon modes or defect density may play a role in the enhanced cooling of G/C-SiC. These insights advance our understanding of how doping and substrate interactions influence carrier dynamics in graphene and offer strategies for optimizing its performance in electronic and optoelectronic applications.

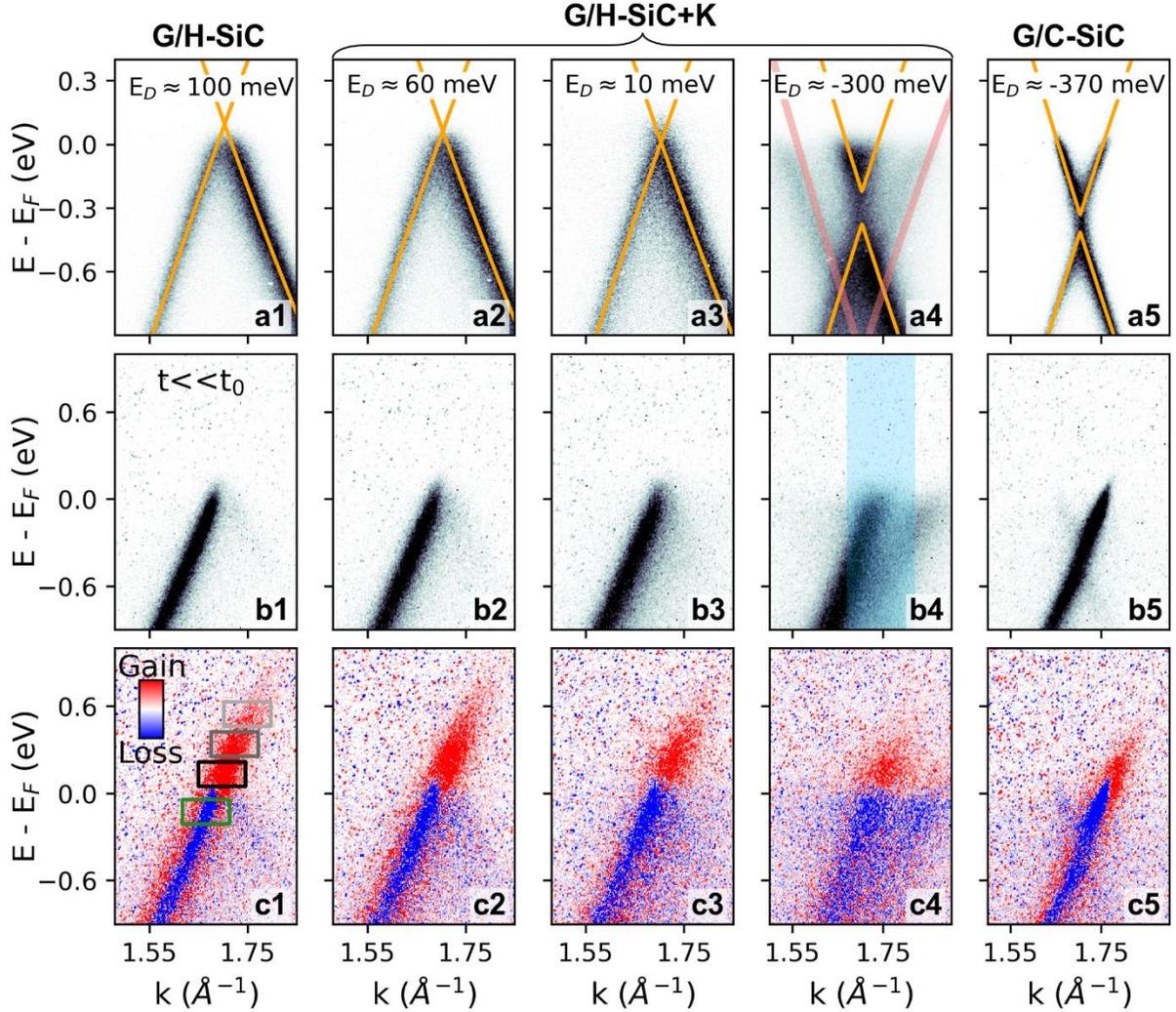

**Figure 1.** Equilibrium and time-resolved ARPES data of graphene for different doping levels. Row a: Equilibrium ARPES images at the K point along the ΓK direction of pristine G/H-SiC (a1), G/H-SiC+K (a2-a4) and G/C-SiC (a5) taken with a Helium lamp. Orange lines represent the tight binding band structure of graphene. Faint red lines in a4 indicate the formation of strongly n-doped domains [16]. Row b: trARPES snapshots for negative pump-probe delays before the arrival of the pump pulse. Row c: Pump-induced changes of the photoemission current 140fs after photoexcitation at a pump photon energy of $\hbar\omega_{pump}$=1.55eV with a pump fluence of 0.9mJ cm$^{-2}$. Gain and loss of photoemitted electrons are shown in red and blue, respectively. Coloured boxes in c1 indicate the areas of integration for pump-probe traces shown in Figure 3 a).

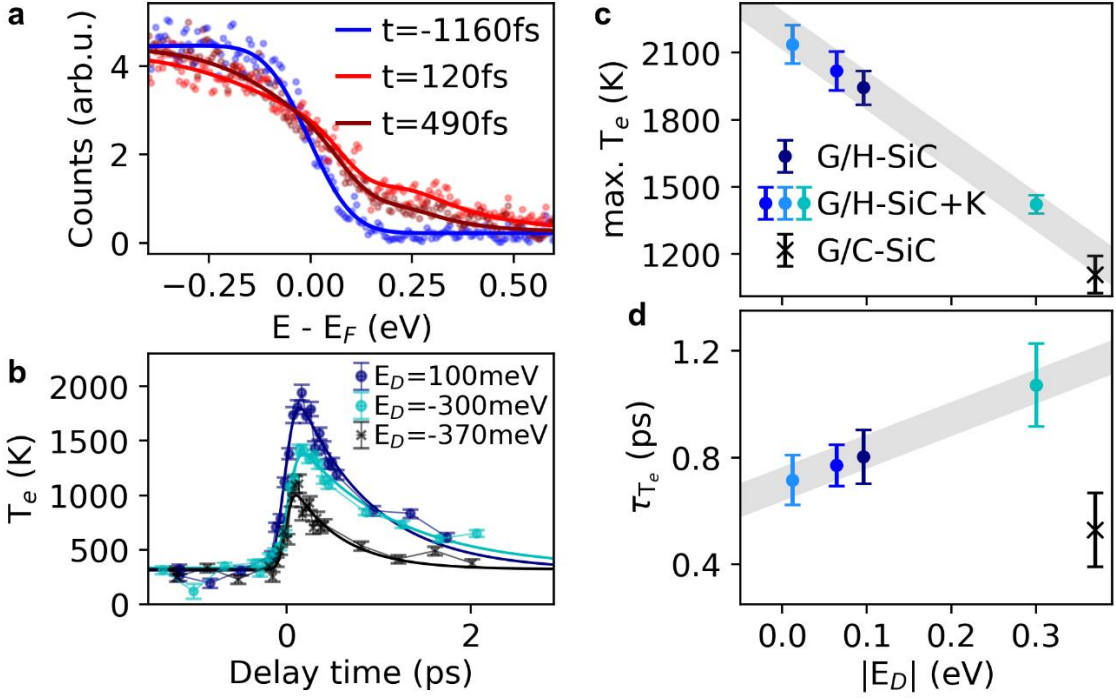

**Figure 2.** Influence of doping level on transient electronic distribution. a) Exemplary energy-distribution curves for different pump-probe delay times obtained by integrating the photocurrent in Fig. 1 b1 over the entire momentum axis. Thick lines are Fermi-Dirac fits. b) Transient electronic temperature for different doping levels together with single-exponential fits. c) Peak electronic temperature for different doping levels corresponding to different energies of the Dirac point $|E_D|$. d) Decay time of the transient electronic temperature for different doping levels. Thick grey lines in c) and d) are guides to the eye.

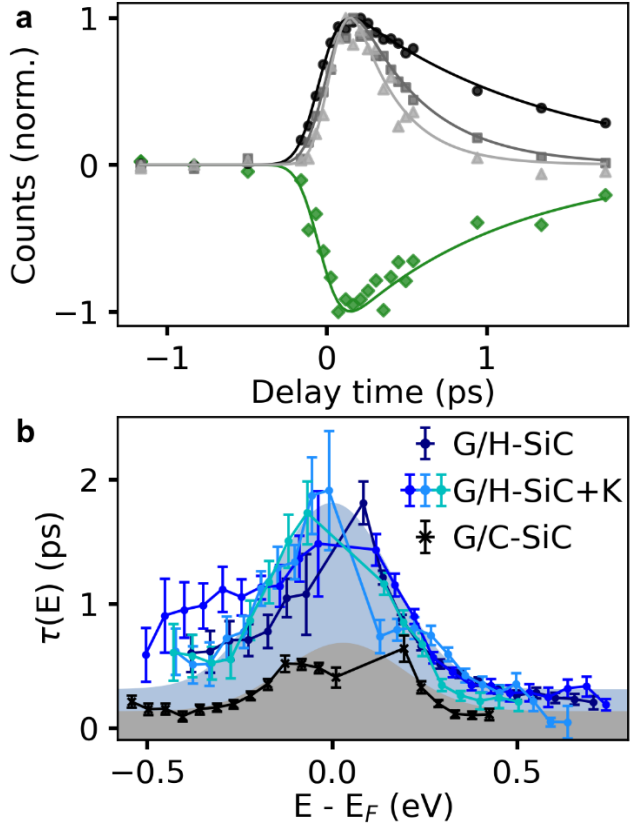

**Figure 3.** Energy-resolved population dynamics for different doping levels. a) Exemplary pump-probe traces obtained by integrating the photocurrent over the areas marked by the coloured boxes in Fig 1 c1. Solid lines are single-exponential fits. b) Energy dependent decay times, τ(E). Blue and grey shaded areas are guides to the eye and differentiate between graphene on H- and C-terminated substrates. The colour of the data points is the same as in Figs. 2c and d.

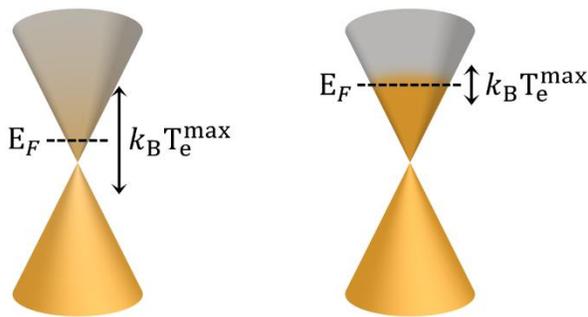

**Figure 4.** Sketch of scattering phase space for two different doping levels. Increased doping comes along with reduced peak electronic temperatures, such that it is not a priori clear if the phase space for phonon emission increases or decreases with increasing carrier concentration.

## ASSOCIATED CONTENT

**Supporting Information**. None

## AUTHOR INFORMATION

**Corresponding Author**

*isabella.gierz@ur.de

**Author Contributions**

LW and JG performed the trARPES experiments, LW analyzed and interpreted the data, PR and TS provided G/H-SiC samples, DC, SF and CC provided G/C-SiC samples, IG conceived the experiment, interpreted the data and wrote the initial draft of the manuscript. All authors have given approval to the final version of the manuscript.

### Acknowledgements

This work received funding from the European Union's Horizon 2020 research and innovation program under Grant Agreement No. 851280-ERC-2019-STG as well as from the German Science Foundation (DFG) via the Collaborative Research Centre CRC 1277 (Project No. 314695032) and the Research Unit RU 5242 (Project No. 449119662).